\documentclass[11pt]{article}
\usepackage[T1]{fontenc}
\usepackage{babel}
\usepackage{graphicx}
\usepackage{color}
\usepackage{bm}
\usepackage{longtable}
\usepackage{amsmath} 
\usepackage{amsfonts}
\usepackage{amssymb}
\usepackage{hyperref}
\DeclareUnicodeCharacter{0301}{*************************************}
\usepackage[sorting=none, style=nature, maxbibnames=99]{biblatex}
\DeclareFieldInputHandler{title}{}
\usepackage[paperwidth=595.44pt,paperheight=841.68pt,top=72pt,right=72pt,bottom=72pt,left=72pt]{geometry}
\begin{document}
\title{\bf Condition on n-Qubit State For Getting Perfect Quantum Teleportation\footnote{This paper is dedicated in the memory of Late Prof. R. Prakash and Late Prof. H. Prakash}}
\author{ Shamiya Javed\footnote{javedshamiya@allduniv.ac.in}, Phool Singh Yadav\footnote{phsyadav@rediffmail.com}, Ranjana Prakash and Hari Prakash\\
Physics Department, University of Allahabad, Prayagraj, India}
\date{}
\maketitle
\begin{abstract}
It is shown that standard quantum teleportation (SQT) with multi-qubit resource result in fidelity $(2+C)/3$ where $C$ is concurrence of the resource in bipartite entanglement between qubit going to receiver and rest of the qubits. For perfect SQT, obviously, $C=1$. For a general 3-qubit resource, we find conditions for getting perfect SQT for state expressed in any basis states. Zha \textit{et al} [Mod. Phys. Lett. B 22, 2523-2528 (2008)], who studied perfect SQT using 3-qubit resource, reported conditions for perfect SQT for only those resource states which are given in the 3-qubit canonical form of Acin representation. We show that there is an alternative easily derivable representation which gives more generalized results. To illustrate the difference between the two schemes, we build an example of 3-qubit entangled state, giving perfect SQT and not included in Zha \textit{et al} results.
\end{abstract}
 \section{Introduction}
\noindent Quantum entanglement stands as the strange phenomenon of quantum mechanics which was first predicted by Einstein in his famous paper \cite{PhysRev.47.777}. Since then it has been studied by several authors \cite{riedinger2018remote,lin2020quantum}. It has many applications in quantum information theory such as quantum teleportation \cite{PhysRevLett.70.1895}, quantum cryptography \cite{bennett1992quantum}, superdense coding \cite{bennett1992communication}, quantum computation \cite{divincenzo1995quantum}, quantum secure direct communication \cite{qi202115,sheng2022one,zhou2022one}, quantum machine learning \cite{sheng2017distributed} and so on. Quantum teleportation (QT) is one of the most important applications and was first proposed by Bennet {\it et al.} \cite{PhysRevLett.70.1895}. In standard quantum teleportation (SQT), the information of a quantum two level system is transmitted from a sender (Alice) to a far apart receiver (Bob) using a 2-qubit maximally entangled (ME) state and a 2-bit classical communication channel. Afterwards, many researchers made significant advances in theory \cite{karlsson1998quantum,shi2002teleportation,joo2003quantum,prakash2011quantum,
ren2017ground,fatahi2021quantum} and experiment \cite{bouwmeester1997experimental,hu2020experimental} with QT. Many recent progress in the direction of QT has also been made such as, long distance QT \cite{ren2017ground}, high dimensional QT \cite{hu2020experimental}, logic qubit QT   \cite{zhou2015complete,sheng2015two} and bidirectional 
QT \cite{wang2015quantum,mastriani2021bidirectional,pandey2019,prakash2019controlled}. Due to the tight connection between quantum entanglement and QT, numerous studies have been conducted on the use of multi-particle entangled quantum states other than two-particle entangled states for QT. In particular, 3-qubit entangled GHZ state\cite{karlsson1998quantum} and 3-qubit entangled W state\cite{shi2002teleportation,joo2003quantum} have been explored. The QT involving ME states (concurrence \cite{wootters1998entanglement}, $C=1$) achieves with unit fidelity and unit success probability, which is termed as perfect QT. The perfect QT of a single qubit via 3-qubit entangled GHZ state \cite{karlsson1998quantum} and  3-qubit entangled GHZ like state \cite{prakash2011quantum} is well studied. However, using a 3-qubit W state \cite{shi2002teleportation,joo2003quantum} perfect teleportation with non-unit success probability is obtained, which is called probabilistic QT. Since in real scenario the ME state interacts with the environment resulting in the decoherence\cite{PhysRevA.64.022313,prakash2008effect} of the ME state and changes it to a non-maximally entangled (NME) state. Therefore, several authors \cite{PhysRevA.61.034301,AGRAWAL200212,yan2010,javed2021high} have also investigated the probabilistic QT of a single qubit using various NME states.

Agrawal and Pati \cite{agrawal2006perfect} proposed that a class of 3-qubit W-state resource leads to perfect SQT. Further, Li {\it et al.} \cite{li2007states} generalized the results of Agrawal and Pati \cite{agrawal2006perfect} to higher dimensional quantum systems. Recently, much attentions have been taken to this
 topic \cite{zuo2009simpler,singh2016usefulness,zha2008two,
 shang2009necessary,tan2016perfect,
muralidharan2008perfect} and various
 schemes \cite{li2016generating,ozaydin2021deterministic} have been proposed for generating such states giving perfect QT. Zuo {\it et al.} \cite{zuo2009simpler} gave a criterion on W-states being suitable for perfect teleportation and superdense coding. A multi-qubit W-type state  has also been suggested by Singh
  {\it et al.} \cite{singh2016usefulness}to accomplish this task. Zha {\it et al.} \cite{zha2008two} gives two forms of 3-qubit entangled states and reported that perfect SQT can be obtained using only those general 3-qubit states which can be expressed in one of their two forms. For an n-qudit entangled resource, Li {\it et al.} \cite{shang2009necessary} suggested that in order to provide perfect SQT the state must satisfy the total $\frac12(d+2)(d-1)$ constraints. In all these schemes various entangled resource states have been suggested for perfectly teleporting an arbitrary single qubit state. However, no one gives a general form of n-qubit entangled resource that will leads to perfect SQT. It should be interesting to find conditions on n-qubit entangled state, so that one can use it as a resource to teleport a single qubit state perfectly.

We study SQT of a single qubit using a general n-qubit resource state expressed in any arbitrary basis. We provide a simple scheme to find maximal average fidelity (MAF) in this general case and conditions for perfect SQT. This result has been used to obtain conditions for perfect SQT using a general 3-qubit resource given in any basis. Zha {\it et al} \cite{zha2008two} have also given results for perfect SQT with 3-qubit states. However, their results are usable only if the 3-qubit resource state is expressed in one particular Acin representation \cite{acin2000generalized,acin2001three}.
\begin{figure}[h!]
\centering
\includegraphics[scale=0.8]{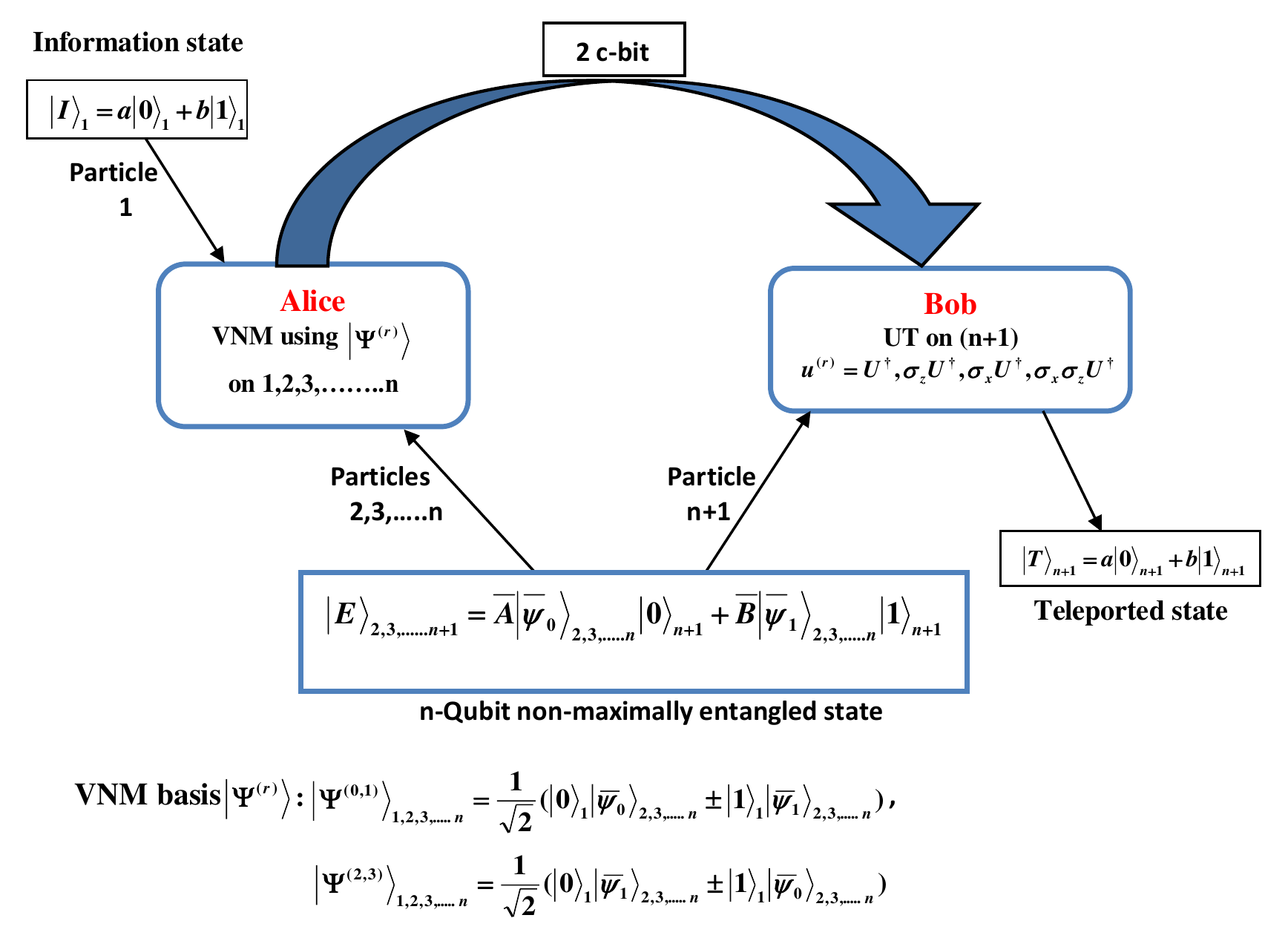}
\caption{Schematic diagram of the proposed scheme of perfect SQT via n-qubit entangled resource.}\label{fig}
\end{figure}

\section{Perfect SQT scheme using a multi-qubit entangled resource}

The sketch diagram of the SQT scheme using a general n-qubit resource is shown in Fig. \ref{fig}. We consider an arbitrary n-qubit non-maximally entangled resource, $|E\rangle_{2,3,.....,n,n+1}$, with n-1 qubits (2,3,4,.....n) go to Alice and one qubit (n+1) goes to Bob. First Alice makes an n-qubit projective measurement or von Neumann measurement (VNM) in the orthogonal basis $\{|\Psi^{(r)}\rangle\}$ on information qubit and her shared entangled qubits. Then she sends her measurement result $r=0,1,2,3$ via 2 bit classical channel to Bob, who makes a suitable unitary transformation (UT), $u^{(r)}$ on his qubit to recover the information state. Let Alice has an unknown information state, 
\begin{equation}
|{I}\rangle_{1}=a|{0}\rangle_{1}+b|{1}\rangle_{1}\label{eq1}
\end{equation}
with the normalization $|a|^2+|b|^2=1$, required to be teleported to a far distant receiver Bob. Also, consider n-qubit entangled resource of which n-1 qubits (numbered as 2,3,4....n) go to Alice and one qubit (numbered as n+1) goes to Bob. We separate Bob's qubit and write resource state as 
\begin{equation}
|{E}\rangle_{2,3,....n+1}=A|{\psi_{0}}\rangle_{2,3,....n}|{0}\rangle_{n+1}+B|{\psi_{1}}\rangle_{2,3,....n}|{1}\rangle_{n+1}\label{eq2}
\end{equation} 
where $|{\psi_{0}}\rangle$ and $|{\psi_{1}}\rangle$ are two states of n-1 qubits going to Alice, which may or may not be orthogonal or entangled.
For qubit n+1, let us use new basis states $|{\bar{0}}\rangle=U|{0}\rangle\equiv(1+|z|^{2})^{-1/2}(|{0}\rangle+z|{1}\rangle)$ and $|{\bar{1}}\rangle=U|{1}\rangle\equiv(1+|z|^{2})^{-1/2}(|{1}\rangle-z^*|{0}\rangle)$ respectively, where $z$ is a complex number, and write the Eq.(\ref{eq2}) as,
\begin{eqnarray}
\nonumber|{E}\rangle_{2,3,....n+1}&=&\frac{1}{1+|z|^2}[A|{\psi_{0}}\rangle_{2,3,....n}\{|{\bar{0}}\rangle_{n+1}-z|{\bar{1}}\rangle_{n+1}\}+B|{\psi_{1}}\rangle_{2,3,....n}\{z^*|{\bar{0}}\rangle_{n+1}+|{\bar{1}}\rangle_{n+1}\}],\\\nonumber
&=&\frac{1}{1+|z|^2}[A|{\psi_{0}}\rangle_{2,3,....n}+Bz^*|{\psi_{1}}\rangle_{2,3,....n}]|{\bar{0}}\rangle\\
&&+\frac{1}{1+|z|^2}[-Az|{\psi_{0}}\rangle_{2,3,....n}+B|{\psi_{1}}\rangle_{2,3,....n}]|{\bar{1}}\rangle.
\end{eqnarray}
 This enables us to write the entangled resource (\ref{eq2}) in the form 
\begin{equation}
|{E}\rangle_{2,3,....n+1}=\bar{A}|{\bar{\psi_{0}}}\rangle_{2,3,....n}|{\bar{0}}\rangle_{n+1}+\bar{B}|{\bar{\psi_{1}}}\rangle_{2,3,....n}|{\bar{1}}\rangle_{n+1}\label{eq3}
\end{equation}
where $|{\bar{\psi_{0}}}\rangle$ and $|{\bar{\psi_{1}}}\rangle$ are orthogonal and $\bar{A}$ and $\bar{B}$ are real and positive, $z$ is then a solution of 
\begin{equation}
Kz^{2}+(A^{2}-B^{2})z-K^*=0,\label{eq4}
\end{equation} 
with 
\begin{equation}
K=\langle{\psi_{1}}|{\psi_{0}}\rangle, \label{eq5}
\end{equation}
 \begin{equation}\bar{A}=(1+|z|^2)^{-1/2}[A^2+B^2|z|^2+AB(Kz+K^*z^*)]^{-1/2},\label{eq6}
 \end{equation}
 \begin{equation}
  \bar{B}=(1+|z|^2)^{-1/2}[B^2+A^2|z|^2-AB(Kz+K^*z^*)]^{-1/2}.\label{eq7}
  \end{equation}
This state given by Eq.(\ref{eq3}) has bipartite entanglement between (n-1) qubits of Alice and 1 qubit of Bob with concurrence $C=2\bar{A}\bar{B}$. It should be noted that in the calculation of bipartite concurrence between sender and receiver we considered all the (n-1) qubits of Alice as a single qubit.\\
If Alice makes projective measurements on qubits 1,2,....n on four mutually orthogonal states 
\begin{eqnarray}
\nonumber|{\Psi^{(0,1)}}\rangle &=&\frac{1}{\sqrt{2}}[|{0}\rangle_{1}|{\bar{\psi_{0}}}\rangle_{2,....n}\pm|{1}\rangle_{1}|{\bar{\psi_{1}}}\rangle_{2,....n}],\\
 |{\Psi^{(2,3)}}\rangle &=&\frac{1}{\sqrt{2}}[|{0}\rangle_{1}|{\bar{\psi_{1}}}\rangle_{2,....n}\pm|{1}\rangle_{1}|{\bar{\psi_{0}}}\rangle_{2,....n}]\label{eq8}
\end{eqnarray}
 the probabilities $P^{(r)}$ for detection of states $|{\Psi^{(r)}}\rangle$, where $r=0,1,2,3$ are given by 
\begin{eqnarray}
\nonumber P^{(0)}&=&P^{(1)}=\frac{1}{2}[|a\bar{A}|^2+|b\bar{B}|^2],\\
 P^{(2)}&=&P^{(3)}=\frac{1}{2}[|b\bar{A}|^2+|a\bar{B}|^2]
\end{eqnarray}
 respectively. The corresponding Bob's states are 
 \begin{eqnarray}
 \nonumber|Bob^{(0)}\rangle &=&\frac{1}{\sqrt{2P^{(0)}}}[a\bar{A}|{\bar{0}}\rangle+ b\bar{B}|{\bar{1}}\rangle],\\\nonumber
  |Bob^{(1)}\rangle &=&\frac{1}{\sqrt{2P^{(1)}}}[a\bar{A}|{\bar{0}}\rangle- b\bar{B}|{\bar{1}}\rangle],\\ 
 \nonumber|Bob^{(2)}\rangle &=&\frac{1}{\sqrt{2P^{(2)}}}[b\bar{A}|{\bar{0}}\rangle+ a\bar{B}|{\bar{1}}\rangle],\\
  |Bob^{(3)}\rangle &=&\frac{1}{\sqrt{2P^{(3)}}}[b\bar{A}|{\bar{0}}\rangle- a\bar{B}|{\bar{1}}\rangle] \label{eq11}
 \end{eqnarray} 
 respectively. If Bob then makes unitary transformations $U^{\dagger}$, $\sigma_{z}U^{\dagger}$, $\sigma_{x}U^{\dagger}$, $\sigma_{x}\sigma_{z}U^{\dagger}$ respectively, SQT with fidelities 
 \begin{eqnarray}
\nonumber F^{(0)}&=& F^{(1)}=\frac{1}{2P^{(0)}}||a|^2\bar{A}+|b|^2\bar{B}|^2,\\
  F^{(2)}&=& F^{(3)}=\frac{1}{2P^{(2)}}||b|^2\bar{A}+|a|^2\bar{B}|^2 \label{eq12}
 \end{eqnarray}
  respectively are obtained. On using average values $\langle|\bar{a}|^4\rangle=\langle|\bar{b}|^4\rangle=1/3$ and $\langle|ab|^2\rangle=1/6$ for randomly changing values of $a$ and $b$, we obtain maximal average fidelity (MAF) as, 
  \begin{equation}
 MAF=<\sum_{r} P^{(r)}F^{(r)}>=\frac{(2+C)}{3}\label{eq13}
 \end{equation}
where $C=2\bar{A}\bar{B}$, concurrence of the resource in bipartite entanglement between qubit going to receiver and rest of the qubits. For perfect SQT,  the right hand side of Eq.(\ref{eq13}) should be $1$, which means $C$ should be $1$. To fulfill this, we demand $\bar{A}=\bar{B}=1/\sqrt{2}$, which gives $A=B=1/\sqrt{2}$ and $\langle{\psi_{1}}|{\psi_{0}}\rangle=0$ also. This simply implies that to achieve perfect SQT, the resource state should have perfect bipartite entanglement between qubit going to Bob and rest of the qubits.

\section{Perfect SQT scheme using a 3-qubit entangled resource}
The most general 3-qubit state in any basis can be written as 
\begin{equation}
|{\Psi}\rangle_{123}=A|{000}\rangle+B|{010}\rangle+C|{100}\rangle+D|{110}\rangle+E|{001}\rangle+F|{011}\rangle+G|{101}\rangle+H|{111}\rangle\label{eq14}
\end{equation}
with the normalization condition $|A|^{2}+|B|^{2}+|C|^{2}+|D|^{2}+|E|^{2}+|F|^{2}+|G|^{2}+|H|^{2}=1$.\\
This is written in the form of Eq.(\ref{eq2}), equality of two constants and the condition $\langle{\psi_{1}}|{\psi_{0}}\rangle=0$ gives    
\begin{equation}
|A|^{2}+|B|^{2}+|C|^{2}+|D|^{2}-|E|^{2}-|F|^{2}-|G|^{2}-|H|^{2}=0,\; A^*E+B^*F+C^*G+D^*H=0\label{eq15}
\end{equation}   
as the conditions for getting perfect SQT with qubit 3 going to Bob, in any measurement basis.\\
In order to write a 3-qubit state satisfying both the conditions of Eq.(\ref{eq15}), we separate Bob's qubit and write a general 3-qubit resource state as,
\begin{equation}
|{\Psi}\rangle_{123}=A_0|{\psi_{0}}\rangle_{12}|{0}\rangle_{3}+A_1|{\psi_{1}}\rangle_{12}|{1}\rangle_{3}.\label{eq16}
\end{equation} 
Here $|{\psi_{1}}\rangle$ may be separable or entangled. If it is separable, one can cast $|{\psi_{1}}\rangle$  in the form $|11\rangle_{12}$ with proper choice of states $|1\rangle_{1}$ and $|1\rangle_{2}$. Orthogonal state $|{\psi_{0}}\rangle_{12}$ then involves states  $|00\rangle_{12}$, $|01\rangle_{12}$, $|10\rangle_{12}$ and $|\Psi\rangle_{123}$  can be written in the general form by choosing suitable phases of states as,
\begin{eqnarray}
|\Psi\rangle_{123}=[a|{00}\rangle+b|{01}\rangle+\sqrt{\frac{1}{2}-a^2-b^2}|{10}\rangle]_{12}|0\rangle_{3}+\frac{1}{\sqrt{2}}|{111}\rangle_{123}.\label{eq17}
\end{eqnarray}  
If  $|{\psi_{1}}\rangle$ is not separable it can be expressed in 2-qubit Schmidt representation \cite{ekert1995entangled} in the form 
\begin{eqnarray}
|{\psi_{1}}\rangle_{12}=a|00\rangle+\sqrt{1-a^2}|11\rangle\label{eq18}
\end{eqnarray}
 with properly chosen states $|0\rangle_i$  and $|1\rangle_i$,   ($i=1,2$) and real coefficient $a$.
The most general form of orthogonal  $|{\psi_{0}}\rangle_{12}$ can then be written as   
\begin{eqnarray}
|{\psi_{0}}\rangle_{12}=\kappa(\sqrt{1-a^2}|00\rangle-a|11\rangle)+be^{i\beta}|01\rangle+\sqrt{1-\kappa^2-b^2}|10\rangle.\label{eq19}
\end{eqnarray}
Here $a,b,\beta,\kappa$ are real. Then $|\Psi\rangle_{123}$ can be written in the form
\begin{eqnarray}
\nonumber
|{\Psi}\rangle_{123} &=&\frac{1}{\sqrt{2}}[\{\kappa(\sqrt{1-a^2}|00\rangle_{12}-a|11\rangle_{12})+be^{i\beta}|01\rangle_{12}\\
&&+\sqrt{1-\kappa^2-b^2}|10\rangle_{12}\}|0\rangle_3
+\{a|00\rangle_{12}+\sqrt{1-a^2}|11\rangle_{12}\}|1\rangle_3].\label{eq20}
\end{eqnarray}
This is more generalized form of the result suggested by Zha \textit{et al} \cite{zha2008two}, for 3 qubit states available in the Acin representation \cite{acin2000generalized},
\begin{equation}
|{\Psi}\rangle_{123}=\kappa_{0}e^{i\theta}|{000}\rangle+\kappa_{1}|{001}\rangle+\kappa_{2}|{010}\rangle+\kappa_{3}|{100}\rangle+\kappa_{4}|{111}\rangle\label{eq21}
\end{equation}
like the simplifying Schmidt representation \cite{ekert1995entangled} for 2 qubit states  given in terms of only one real and positive parameter and two terms, the Acin \textit{et al} representation expresses 3-qubit state in terms of only five terms which involves only four independent real and positive parameters (four of real and positive parameters $\kappa_{0}$, $\kappa_{1}$, $\kappa_{2}$, $\kappa_{3}$, $\kappa_{4}$) and an angle $\theta$. Zha \textit{et al} reported that perfect SQT will result if $|{\psi}\rangle_{123}$ is in one of the following forms:
\begin{equation}
|{\psi}\rangle_{123}=\kappa_{0}e^{i\theta}|{000}\rangle+\kappa_{2}|{010}\rangle+\sqrt{\frac{1}{2}-\kappa^2_{0}-\kappa^2_{2}}|{100}\rangle+\frac{1}{\sqrt{2}}|{111}\rangle,\label{eq22}
\end{equation}
\begin{equation}
|{\psi}\rangle_{123}=\kappa_{1}|{001}\rangle+\kappa_{2}|{010}\rangle+\sqrt{\frac{1}{2}-\kappa^2_{2}}|{100}\rangle+\sqrt{\frac{1}{2}-\kappa^2_{1}}|{111}\rangle.\label{eq23}
\end{equation}
This result is obvious, if one writes Eq.(\ref{eq21}) in the form of Eq.(\ref{eq2}) and then applies orthogonality condition $\langle{\psi_{1}}|{\psi_{0}}\rangle=0$, which gives $\kappa_{0}\kappa_{1}=0$, resulting in  $|{\psi}\rangle_{123}$ in the form of Eq.(\ref{eq22}) or Eq.(\ref{eq23}), according as $\kappa_{0}=0$ is taken or $\kappa_{1}=0$ is taken.

Here, this must also be noted that one could use the alternative Acin representation \cite{acin2001three} of 3 qubit states, which also involves five real and positive parameters (four of $a, b, c, d, f$ and an angle $\theta$) and only five terms, viz,
\begin{equation}
|{\psi}\rangle_{123}=a|{000}\rangle+b|{100}\rangle+c|{101}\rangle+d|{110}\rangle+fe^{i\theta}|{111}\rangle\label{eq24}
\end{equation}
For this the orthogonality condition $\langle{\psi_{1}}|{\psi_{0}}\rangle=0$ gives $df=0$ as the condition for resource $|{\psi}\rangle_{123}$ to result in perfect SQT. Hence the two permitted forms of $|{\psi}\rangle_{123}$ parallel to Eqs. (\ref{eq22}) and (\ref{eq23}) should contain the first three terms of right hand side of Eq. (24) and any one of the last two. Using our scheme it can easily be shown that there exist 3-qubit entangled states which can lead to perfect SQT and not included in earlier results of Zha \textit{et al}.\\
For example, let us consider a 3-qubit entangled state, 
\begin{equation}
|{E}\rangle_{123}=\frac{1}{\sqrt{2}}e^{i\theta}|000\rangle_{123}+a|011\rangle_{123}+be^{i\delta}|101\rangle_{123}+\sqrt{\frac{1}{2}-a^2-b^2}e^{i\gamma}|111\rangle_{123}\label{eq25}
\end{equation}
where particles 1, 2 belongs to Alice and particle 3 goes to Bob. It can also be written in the form of Eq.(\ref{eq2}) as,
\begin{equation}
|{E}\rangle_{123}=\frac{1}{\sqrt{2}}e^{i\theta}|00\rangle_{12}|0\rangle_3+[a|01\rangle+be^{i\delta}|10\rangle+\sqrt{\frac{1}{2}-a^2-b^2}e^{i\gamma}|11\rangle]_{12}|1\rangle_3.\label{eq26}
\end{equation}
Here $|{\psi_0}\rangle=|00\rangle$ and $|{\psi_1}\rangle=\sqrt{2}[a|01\rangle+be^{i\delta}|10\rangle+\sqrt{\frac{1}{2}-a^2-b^2}e^{i\gamma}|11\rangle]$ implying $\langle{\psi_0}|{\psi_1}\rangle=0$. Also, this state is satisfying the second condition i.e. unit bipartite entanglement between sender's and receiver's qubits. Hence, it can be used to teleport a single qubit perfectly.

Our results also shows directly that the only W-state which can give perfect SQT is \cite{agrawal2006perfect}
 \begin{equation}
|{\psi}\rangle_{123}=B|{100}\rangle+C|{010}\rangle+E|{001}\rangle\label{eq27}
\end{equation} with $|B|^2+|C|^2=|E|^2=\frac{1}{2}$, which also includes the normalization i.e. $|B|^2+|C|^2+|E|^2=1$.  
\section{Conclusions}
We study perfect SQT of a single qubit and find conditions on a most general form of n-qubit resource to give perfect SQT. To achieve this, we consider an n-qubit general entangled resource, where n-1 qubits go to Alice and 1 qubit goes to Bob. Alice makes an n-qubit projective measurement on her particles along with the information particle and sends 2-c bit message to Bob. Then Bob applies a suitable unitary transformation on his particle to replicate the information. We obtained maximal average fidelity for a general n-qubit resource. For perfect SQT, there should be unit bipartite entanglement between qubit belongs to the receiver and rest of the qubits. Using this result we discussed a 3-qubit entangled state, written in a derivable computational basis and giving perfet SQT. This reslut is different from earlier, where entangled resource is written in Acin representation. We have also given an example of 3-qubit entangled state which can give perfect SQT and not included in previous studies. Additionally, we demonstrate how our findings clearly show the existence of a class of 3-qubit W-states that can result in perfect SQT.
\section*{Acknowledgement}
The authors are gratefully acknowledge Prof. Naresh Chandra, Dr. Onkar Nath Verma, Dr. Vikram Verma  and Mr. Ravi Kamal Pandey for their valuable and helpful discussions. One of the authors SJ is also thankful to University Grant Commission scheme of Maulana Azad National Fellowship for providing financial support.
\printbibliography

\end{document}